# Strong perpendicular magnetic anisotropy energy density at Fe alloy/HfO$_2$ interfaces

Yongxi Ou[1*], D. C. Ralph[1,2], and R. A. Buhrman[1+]

[1]Cornell University, Ithaca, New York 14853, USA

[2]Kavli Institute at Cornell, Ithaca, New York 14853, USA

*yo84@cornell.edu,   +rab8@cornell.edu

## ABSTRACT

We report on the perpendicular magnetic anisotropy (PMA) behavior of heavy metal (HM)/ Fe alloy/MgO thin film heterostructures after an ultrathin HfO$_2$ passivation layer is inserted between the Fe alloy and the MgO. This is accomplished by depositing one to two atomic layers of Hf onto the Fe alloy before the subsequent rf sputter deposition of the MgO layer. This Hf layer is fully oxidized during the subsequent deposition of the MgO layer, as confirmed by X-ray photoelectron spectroscopy measurements. As the result a strong interfacial perpendicular anisotropy energy density can be achieved without any post-fabrication annealing treatment, for example 1.7 erg/cm$^2$ for the Ta/Fe$_{60}$Co$_{20}$B$_{20}$/HfO$_2$/MgO heterostructure. Depending on the HM, further enhancements of the PMA can be realized by thermal annealing to at least $400\,^{\circ}C$. We show that ultra-thin HfO$_2$ layers offer a range of options for enhancing the magnetic properties of magnetic heterostructures for spintronics applications.

The realization of robust perpendicular magnetic anisotropy (PMA) in heavy metal (HM)/Fe alloy/MgO thin-film heterostructures[1,2], where typically the Fe alloy is $Fe_{80-x}Co_xB_{20}$ (FeCoB), has enabled a pathway for the implementation of high density memory elements based on the spin transfer torque switching of perpendicularly magnetized tunnel junctions (MTJs)[2–4]. Strong PMA is also required to create the perpendicularly magnetized nanowire structures needed to enable manipulation of domain walls and novel magnetic chiral structures such as skyrmions by the spin Hall effect[5–8]. At present the only viable FM/oxide combination that yields the strong PMA and low damping required for practical devices is $Fe_{80-x}Co_xB_{20}$ (FeCoB)/MgO where the PMA originates from the strong spin-orbit interaction in the hybridized 3d Fe-2p O bonding at the FeCoB/MgO interface[9,10]. Even there obtaining significant PMA requires an annealing step[1–4] that can compromise the layers in the magnetic heterostructure.

In previous research, we studied the modification of field-like spin orbit torque at the FeCoB/MgO interface via tuning the PMA there by introducing an ultra thin Hf oxide layer[11]. Here we report a systematic study that this addition to the surface of FeCoB of as little as 0.2 nm of Hf "dusting", which is oxidized to HfO2 during the subsequent MgO deposition process, can yield strong PMA without any post-fabrication annealing treatment. Depending on the HM, the system can also, if that is desired, be annealed to at least $400\ ^oC$ to further enhance the PMA. This simple Hf dusting technique not only improves the performance of FeCoB/MgO structures but also allows for the PMA devices to be made from the low-damping, low-magnetostriction alloy permalloy ($Ni_{80}Fe_{20}$) and likely other Fe alloys. The technique therefore substantially expands the options for engineering magnetic thin film heterostructures for spintronics.

All the samples in this paper were prepared via standard direct current (DC) sputtering (with RF magnetron sputtering for the MgO layer), with a base pressure $<4\times10^{-8}$ Torr . The DC sputtering condition was 2mTorr Ar pressure, 30 watts power. To form the interfacial $HfO_2$ an ultrathin Hf dusting layer was first sputtered on the FeCoB with a low deposition rate of 0.01 nm/s, and the MgO layer was then sputtered on the Hf layer with a growth rate of 0.005 nm/s (at 100 watts power, 2 mTorr Ar), a process that oxidized the Hf. In each case the top Ta film serves as a capping layer to protect the underlayers from degradation due to atmospheric exposure. The heterostructures were further fabricated into $5 \times 60 \, \mu m^2$ Hall bars for measurements. All the samples were baked at $115 \, °C$ for 1 min twice during the standard photolithography fabrication for photoresist treatment.

We will first discuss results obtained from a set of samples (series A) of Si/SiO$_2$/Ta(6)/FeCoB($t_{FeCoB}$)/HfO$_2$($t_{Hf}$)/MgO(2)/Ta(1) with a range of HfO$_2$ thicknesses, and from two control series, one with a Ta dusting layer (series B), Si/SiO$_2$/Ta(6)/FeCoB($t_{FeCoB}$)/TaO$_x$($t_{Ta}$)/MgO(2)/Ta(1) and one without any dusting layer (series C), Si/SiO$_2$/Ta(6)/FeCoB($t_{FeCoB}$)/MgO(2)/Ta(1) (The numbers in parentheses are the thicknesses in nm). Since the complete oxidation of the insulator at the Fe alloy/oxide interface is held to be critical for the formation of PMA in HM/Fe alloy/oxide heterostructures [10], we performed x-ray photoelectron spectroscopy (XPS) measurements on an as-grown Ta(6)/FeCoB(1.2)/HfO$_2$(0.2)/MgO/Ta series A sample. Ion etching was used to remove most of the Ta capping layer before performing XPS. As shown in Fig. 1(a), the HfO$_2$ $4f_{7/2}$ and $4f_{5/2}$ peaks are clearly displayed at 17.1 eV and 18.8 eV, with only a very small sub-oxide peak at ~ 16.0 eV and no evidence for the Hf metallic $4f_{7/2}$ peak at 14.3 eV. To achieve strong PMA it is also required that the Fe alloy not be oxidized beyond the interfacial Fe-O bonds. Fig. 1(b) shows for the same sample the XPS $2p_{3/2}$ peak of Fe at 706.0 eV, which can be well fit with the narrow asymmetric spin-split peak function characteristic of metallic Fe [12]. For a series C sample without the Hf dusting layer, the Fe $2p_{3/2}$ peak is much broader with a high energy tail indicative of substantial oxidation of the surface Fe during the direct deposition of MgO by rf sputtering[13] (Fig. 1(b)). We also examined the Fe XPS signal for a series B sample with Ta as the dusting layer (0.3 nm). That yielded a metallic signal indistinguishable from the Hf case, again indicating protection of the ferromagnetic surface from significant oxidation. However, the magnetic characteristics of these heterostructures are quite different.

In Fig. 2(a) we present the magnetic moment per area as a function of $t_{FeCoB}$ for a set of series A samples (Hf dusting) and also for series B samples (Ta dusting), as measured by vibrating sample magnetometry (VSM). (The average thickness provided for the dusting layer throughout this paper is that of the deposited metal, before oxidation.) The fit to the Hf dusting series (A) gives a saturation magnetization of $M_s = 1260$ emu/cm$^3$ and a very small apparent "dead layer" thickness $t_d \approx 0.1$ nm, both consistent with previous results from as-deposited Ta/FeCoB/MgO structures[1,11,14]). In contrast, the series B samples indicate $t_d \approx 0.8$ nm and a much larger $M_s = 1800$ emu/cm$^3$; results that are comparable to some previous studies of annealed ($\sim 300\ °C$) Ta/FeCoB/MgO samples where the dead layer[15] has been attributed to undesirable diffusion of Ta into the FeCoB, perhaps to the ferromagnet/oxide interface [16]. Thus we tentatively attribute the thick dead layer in series B samples to the intermixing of Ta and FeCoB during the deposition of the Ta dusting layer.

While Ta/FeCoB/MgO structures with a thin FM layer typically only exhibit, at most, a weak perpendicular magnetic anisotropy (PMA) in the as-deposited state[3,14,15], we obtained robust PMA behavior in as-deposited structures with the HfO$_2$ dusting layer. For example in Figure 2(b) we plot the PMA energy density $K_{eff} \equiv H_a M_s / 2$ as a function of the effective thickness $t_{FeCoB}^{eff} = t_{FeCoB} - t_d$ of the FeCoB for the series A Ta/FeCoB($t_{FeCoB}$)/HfO$_2$(0.2)/MgO samples. Here $H_a$ is the perpendicular magnetic anisotropy field as determined from measurement of the anomalous Hall voltage response to an in-plane magnetic field[17,18] and we use the values of $M_s$ determined from the VSM measurements discussed above. Since we can expect, at least approximately, that when $t_{FeCoB}^{eff}$ is sufficiently large $K_{eff} \cdot t_{FeCoB}^{eff} = (K_v - 2\pi M_s^2) \cdot t_{FeCoB}^{eff} + K_s$ where $K_v$ ($K_s$) is the bulk (interfacial) anisotropy energy density, we can use a linear fit to this plot to determine that the surface anisotropy $K_s = 1.74 \pm 0.09$ erg/cm$^2$. For Ta and Hf base layer systems without the Hf dusting, comparable anisotropies can be obtained only via high temperature ($\geq 200\,°C$) annealing[19–21]. In the inset of Fig. 2(b), we also show the effective PMA energy density $K_{eff}$ for the series B samples with a 0.2 nm Ta dusting layer. Notice that here $K_{eff}$ for Ta dusting is an order smaller than for the Hf dusting.

Consistent with the strong $H_a$ of the HfO$_2$ passivated samples, the coercive field $H_c$ of those PMA structures is relatively high, typically $\geq 300$ Oe, in comparison to quite low values $< 20$ Oe) for the Ta dusting samples. Examples of the field switching that is obtained with an external field applied normal to the film surface are provided in Fig. 2(c) for a Ta(6)/FeCoB(1.1)/HfO$_2$(0.2)/MgO/Ta(1) and a Ta(6)/FeCoB(1.1)/TaO$_x$(0.2)/MgO/Ta(1) sample. Since $H_c$ of such PMA samples depends on both the anisotropy field and its uniformity, which together act to set the depinning field for magnetic reversal, further enhancement in $H_c$ should be expected with refinements in the smoothness and uniformity of such heterostructures.

We measured the perpendicular anisotropy fields $H_a$ as a function of $HfO_2$ thicknesses in a different set of samples Ta(6)/FeCoB(0.8)/ $HfO_2$ ($t_{Hf}$)/MgO/Ta with $t_{Hf} \sim 0.2-0.4$ nm, as indicated in Fig. 2(d). For the as-grown samples, $H_a$ increases with $HfO_2$ thickness and grows above 1 T when $t_{Hf} \geq 0.3$ nm. This is likely due to a more completely continuous $HfO_2$ layer being formed at the FeCoB/MgO interface as $t_{Hf}$ is increased over this range and hence a higher Fe-O-Hf hybridized bond density that enhances the interfacial PMA.

Previously, high temperature post-fabrication annealing treatment has been considered to be necessary to the achievement of robust PMA in HM/FeCoB/MgO heterostructures. There are generally two important functions of this annealing process: (i) removal of the over-oxidation of the FeCoB surface that occurs during MgO deposition[13,22] and (ii) promotion of the out-diffusion of the boron from the initially amorphous FeCoB [13,19] to obtain a more ordered, crystalline FeCo/MgO interface. Our results here indicate that the first function is the more important, or alternatively that the Fe-O-Hf hybridized bonds results in a stronger spin-splitting of the orbitals than does the Fe-O-Mg bonds.

Obtaining strong PMA in HM/Fe alloy/Oxide systems without the necessity of thermal annealing may facilitate important applications as this could avoid complications such as material diffusion/intermixing during high temperature excursions. On the other hand, since many applications of PMA heterostructures do require high temperature processing, both for integration with Si circuits and to attain a high tunneling magnetoresistance (TMR) with MTJs, we also studied how different heat treatments affect the PMA of our $HfO_2$ structures. We show in Fig. 2(d) that after annealing at $210\ °C$ for 1 hour, $H_a$ increases for every $HfO_2$ thickness studied, while the general dependence of $H_a$ on $t_{Hf}$ remains. However, after annealing at 300 $°C$ for 1 hour the PMA deteriorates, with a much weaker PMA retained only for $t_{Hf} \geq 0.3$ nm. This deterioration may be due to the diffusion of Ta from the base layer since such diffusion has been known to damage the interfacial PMA in the Ta based PMA systems[3,16].

We have also examined whether this Ta in-diffusion problem can be avoided by the use of other heavy metal base layers, especially those with strong spin Hall effects, e.g. W and Pt. In Fig. 3(a) we show the values of $H_a$ obtained from a set of W(4)/FeCoB(0.8)/HfO$_2$($t_{Hf}$)/MgO(1.6)/Ta samples as a function of $t_{Hf}$ for the as-deposited case, after 1 hour at 300 $°C$, and after 1 hour at 410 $°C$. Here the W is in the high resistivity beta-W phase. The anisotropy increases with annealing temperature, and with 410 $°C$ vacuum annealing we obtain $H_a > 1.6$ T for a sufficiently thick HfO$_2$ passivation layer, indicative of an interfacial anisotropy energy density $\geq 1.5$ ergs/cm$^2$. When we use a 1 nm Ta seeding layer before the deposition of the W layer, it results in the W being smoother and in it also being in the lower resistivity alpha-phase. As shown in Fig. 3(b), relatively high anisotropy fields are obtained after 300 $°C$ annealing of such Ta(1)/W(4)/FeCoB(0.8)/HfO$_2$($t_{Hf}$)/MgO(1.6)/Ta samples for $t_{Hf} \geq 0.1$ nm, but annealing at 410 $°C$ degrades $H_a$, particularly for the heterostructures with thinner HfO$_2$, likely due to in-diffusion of Ta from the bottom seeding layer.

While most PMA heterostructure research currently utilizes either Ta/FeCoB/MgO or Pt/Co/Oxide multilayers, where in the latter case the PMA is concluded to originate largely from spin orbit effects at the Pt/Co interface, other magnetic layers with attractive properties, such as Ni$_{80}$Fe$_{20}$, could be of possible interest and value if samples of sufficiently strong anisotropy can be produced. We find that significant interfacial anisotropy can be obtained with a suitable combination of $HfO_2$ and $Ni_{80}Fe_{20}$, *e.g.* with Ta/ $Ni_{80}Fe_{20}$/$HfO_2$/MgO and with Ta/Hf(0.5)/$Ni_{80}Fe_{20}$/$HfO_2$/MgO multilayers. Our best results to date have been obtained with an amorphous Hf(0.5) spacer between the Ta base layer and the NiFe, which presumably helps to accommodate the crystalline mismatch between the Ta and the NiFe. In Fig. 3(c) we show anomalous Hall measurements as a function of an in-plane magnetic field for as-deposited Ta based NiFe(1.5)/$HfO_2$(0.2)/MgO samples with and without the Hf spacer at the Ta/NiFe interface. $H_a$ for the structure without Hf spacer was 1.1 kOe, while for the sample with the 0.5 nm Hf spacer, $H_a$ is doubled to 2.1 kOe, indicative of an interfacial anisotropy energy density $K_s \approx 0.8 \, \text{erg/cm}^2$, which is somewhat surprising given the low Fe concentration. We have also obtained similar values of $K_s$ with Pt/Hf(0.5)/FeCoB/$HfO_2$(0.2)/MgO multilayers both as deposited and after $300 \, °C$ annealing. We conclude that the combination of a $HfO_2$ passivation layer at the Fe alloy/oxide interface together with a thin Hf spacer layer between the HM and the Fe alloy (when needed due to crystalline mismatch between the HM and the Fe alloy) can be a robust strategy for engineering the PMA of a range of thin-film magnetic heterostructures.

Finally, a recent development in MTJ technology for spin transfer torque applications is to include a second, thinner MgO layer on the other side of the FeCoB free layer, opposite to the MgO tunnel barrier interface[23,24]. This enhances $K_{eff}$ of the free layer permitting the use of a thicker layer with more thermal stability, and also suppresses the magnetic damping enhancement that would otherwise occur via spin pumping to the adjacent normal metal contact. We have examined a modification of this approach by depositing multilayer stacks of MgO(1.6)/FeCoB($t_{FeCoB}$)/$HfO_2$(0.2)/MgO(0.8)/Ta onto oxidized Si substrates. In Fig. 3 (d) we show a plot of $H_a$ of such samples as a function of $t_{FeCoB}$. Quite strong anisotropy fields are obtained to high values of $t_{\text{FeCoB}}$, particularly for the samples annealed at $370\ °C$. Field modulated ferromagnetic resonant studies of such a heterostructure with $t_{\text{FeCoB}} = 1.6$ nm yielded a magnetic damping parameter $\alpha = 0.009$, while a Ta/FeCoB(1.6)/MgO(1.6)/Ta sample showed $\alpha = 0.02$, consistent with earlier work[1,25].

An important question in terms of application is whether MTJ's with a $HfO_2$ passivation layer at the tunnel barrier – free layer interface can provide sufficiently high TMR to useful for STT and other spintronics applications. As reported previously[26], a TMR of 80% has been achieved with an in-plane magnetized Pt/Hf/FeCoB (1.6)/MgO(1.6)/FeCoB/Ru/Ta MTJ structure annealed at 300 C, where analytical STEM reveals substantial Hf within the tunnel barrier and the greatly reduced demagnetization field, $\approx 4$ kOe, indicates a substantial $K_s$. As will be reported elsewhere [27], we are now obtaining similar results with W base layer MTJs, so hybrid $HfO_2$MgO tunnel barriers may well provide sufficient TMR to be of technology interest.

In summary, we have demonstrated that perpendicular magnetic anisotropy in HM/Fe alloy/MgO heterostructures can be dramatically strengthened by incorporating a very thin $HfO_2$ dusting layer at the Fe alloy/MgO interface. In HM/FeCoB/MgO devices, the dusting layer enables strong PMA even in the absence of the post-deposition annealing step that has previously been necessary. When annealing is desired, the dusting layer allows the PMA to remain strong for annealing temperatures even above $400\ °C$, provided a proper base layer is utilized, a much higher limit than for some current STT-MRAM prototype technologies. This can allow easier integration with Si circuitry. The $HfO_2$ dusting can also create robust PMA using magnetic materials for which previously this has been impossible, thereby expanding the portfolio of magnetic materials available for PMA technologies beyond just FeCoB. In particular we have demonstrated PMA with thin-film $Ni_{80}Fe_{20}$, a material that is attractive for its low damping and low magnetostriction. Overall, the strengthening of PMA with the use of $HfO_2$ dusting layers has great promise both for enhancing the performance of spin-transfer-torque magnetic memory based on PMA magnetic tunnel junctions and also for improving control of chiral domain walls and skyrmion structures within PMA HM/Fe alloy/MgO structures[5,6,28–30].

**Acknowledgements**

We thank Darren Dale for the assistance with the XPS measurements and analysis. This research was supported by ONR and by NSF/MRSEC (DMR-1120296) through the Cornell Center for Materials Research (CCMR), and by NSF through use of the Cornell Nanofabrication Facility (CNF)/NINN (ECCS-1542081) and the CCMR facilities.

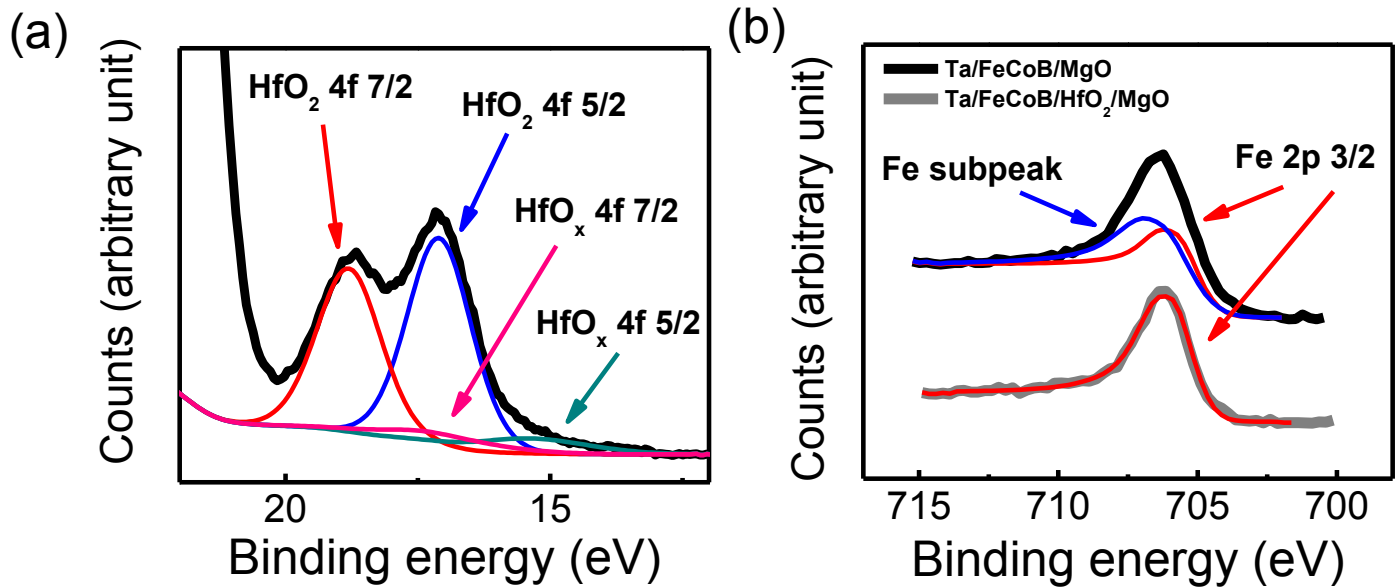

Figure 1.(color online) XPS spectra from (a) HfO$_2$ 4f and (b) Fe 2p spectral regions for the as-grown samples Ta(6)/FeCoB(1.2)/HfO$_2$(0.2)/MgO/Ta(1) and Ta(6)/FeCoB(1.3)/MgO/Ta(1).

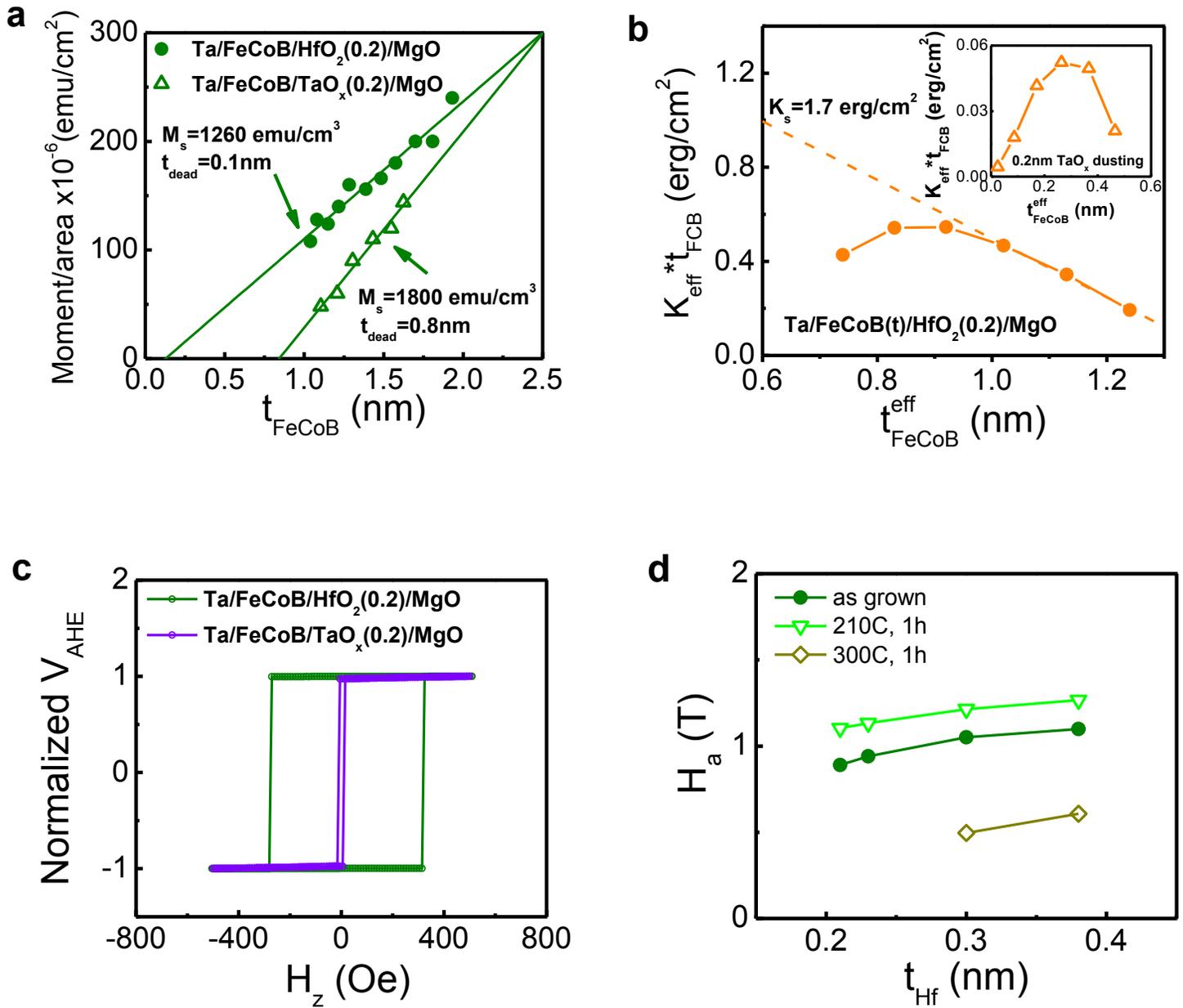

Figure 2.(color online) (a) VSM measurements of magnetization, (b) effective anisotropy energy density $K_{eff}$ determined from anomalous Hall measurements as a function of in-plane magnetic field, and (c) anomalous Hall measurements as a function of out-of-plane magnetic field, for the as-grown samples Ta(6)/FeCoB($t_{FeCoB}$)/HfO$_2$(0.2)/MgO/Ta and Ta(6)/FeCoB($t_{FeCoB}$)/TaO$_x$(0.2)/MgO/Ta. The solid and dashed straight lines are linear fits to the data. (d) The perpendicular anisotropy fields of Ta(6)/FeCoB(0.8)/HfO$_2$($t_{Hf}$)/MgO/Ta samples as deposited and after different post-fabrication annealing treatments.

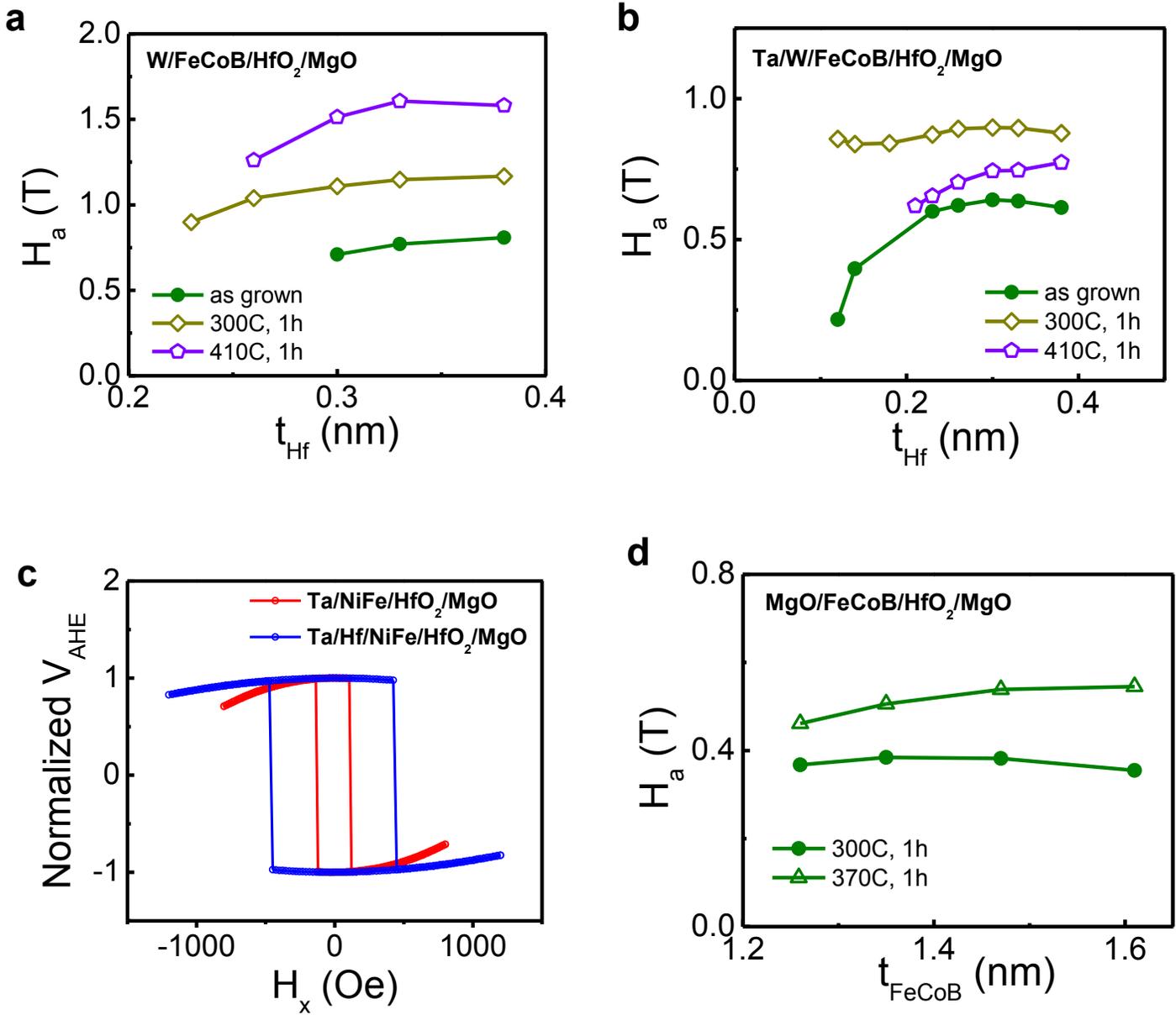

Figure 3.(color online) The perpendicular anisotropy fields of (a) beta-W(4)/FeCoB(0.8)/HfO$_2$($t_{Hf}$)/MgO/Ta and (b) Ta/alpha-W(4)/FeCoB(0.8)/HfO$_2$($t_{Hf}$)/MgO/Ta after different post-fabrication annealing treatments. (c) Anomalous Hall measurements of the as-grown samples Ta(6)/NiFe(1.4)/HfO$_2$(0.2)/MgO/Ta and Ta(6)/Hf(0.5)/NiFe(1.5)/HfO$_2$(0.2)/MgO/Ta as a function of in-plane magnetic field. (d) The perpendicular anisotropy fields of MgO(1.6)/FeCoB($t_{FeCoB}$)/HfO$_2$(0.3)/MgO(0.8)/Ta samples after different post-fabrication annealing treatments.